\documentclass[fleqn,usenatbib]{mnras}

\usepackage{newtxtext,newtxmath}

\usepackage[T1]{fontenc}
\usepackage{ae,aecompl}

\usepackage{graphicx}	
\usepackage{amsmath}	

\title[pressure-mode splitting asymmetry]{Oscillation Frequencies of Moderately Rotating Delta Scuti Stars: Asymmetric Mode Splittings Due to Non-spherical Distortion}

\author[Guo et al.]{
Zhao Guo,$^{1,2,3,4}$\thanks{E-mail: zhao.guo@kuleuven.be},
Timothy R. Bedding$^{1}$, A. A. Pamyatnykh$^{3}$, Donald W. Kurtz$^{5,6}$, Gang Li$^{2}$, 
\and Anuj Gautam$^{7}$,  Simon J. Murphy$^{7}$, Conny Aerts$^{2,8,9}$
\\
${1}$ Sydney Institute for Astronomy (SIfA), School of Physics, University of Sydney, NSW 2006, Australia
\\
${2}$ Institute of Astronomy (IvS), KU Leuven, Celestijnenlaan 200D, 3001 Leuven, Belgium \\
${3}$ Copernicus Astronomical Center, Polish Academy of Sciences, Bartycka 18, Warsaw, Poland \\
${4}$ Department of Applied Mathematics and Theoretical Physics, University of Cambridge, Cambridge CB3 0WA, UK \\
${5}$ Centre for Space Research, North-West University, Dr Albert Luthuli Drive, Mahikeng 2735, South Africa \\
${6}$ Jeremiah Horrocks Institute, University of Central Lancashire, Preston PR1 2HE, UK\\
${7}$ Centre for Astrophysics, University of Southern Queensland, Toowoomba, QLD 4350, Australia
\\
${8}$ Department of Astrophysics, IMAPP, Radboud University Nijmegen, PO Box 9010, 6500 GL Nijmegen, The Netherlands \\
${9}$ Max Planck Institute for Astronomy, Koenigstuhl 17, 69117 Heidelberg, Germany \\
}

\date{Accepted XXX. Received YYY; in original form ZZZ}

\pubyear{2015}

\begin{document}
\label{firstpage}
\pagerange{\pageref{firstpage}--\pageref{lastpage}}
\maketitle

\begin{abstract}
We find that the observed pressure-mode rotational splittings of slowly/moderately rotating $\delta$~Scuti stars and $\beta$~Cephei stars mostly have a positive asymmetry. That is, the left frequency spacing is larger than the right spacing in the dipole mode splitting triplets and the $l=2$ mode splitting multiplets (considering $m=1, 0, -1$ modes only). This is in agreement with the second-order perturbative effect of the rotational non-spherical distortion: both the prograde and retrograde modes have their frequencies shifted towards lower values relative to the $m=0$ modes. We thus study the rotational perturbation both in the first and second order, as well as the near-degeneracy mode coupling effect in MESA models representing $\delta$ Scuti stars. For faster rotators, the near-degeneracy mode coupling between the nearest radial and quadrupole modes can significantly shift the $m=0$ modes, reduce the splitting asymmetry, and even change its sign. We find the theoretical splitting asymmetry from the second-order non-spherical distortion is larger than observed asymmetry. To facilitate future detections, we predict correlations between splitting asymmetry, splitting amplitude, and pulsation frequency. We also discuss additional factors that can influence splitting asymmetry, including embedded magnetic fields, resonant mode coupling, and binarity.


\end{abstract}




\section{Introduction}

The frequencies of oscillation modes in stars are affected by rotation. For slow rotation, the effect is well-described by a simple and symmetric splitting of each mode into a multiplet \citep[e.g.,][]{aer10}.  The situation is more complicated for rapid rotation, which leads to complex mode geometries \citep[see the review by][]{mir22}.  In this paper, we study the rotational splittings of slowly and moderately rotating $\delta$\,Sct stars, for which perturbation theory is still valid. We particularly focus on the splitting asymmetry of the $m=-1, 0, 1$ modes\footnote{We use the convention that $m=-1$ and $1$ correspond to retrograde and prograde modes, respectively.}. 
Observationally, it is customary to interpret the nearly-equally-spaced frequency peaks in the triplets or multiplets as possible rotational splittings. First-order perturbation implies that the splitting is $\approx (1-C_{nl})\Omega_{\rm rot}$, where $C_{nl}$ is the Ledoux constant \citep{led51}, and $\Omega_{\rm rot}$ is the rotational frequency.

Nearly-continuous observations from space missions such as {\it Kepler} and {\it TESS} enable us to measure the oscillation frequencies with high resolution. The precise measurements of rotational splitting asymmetry, which are not usually scrutinised, can provide critical insights into the stellar rotation, magnetism, and mode dynamics.
For example, \citet{li22} inferred an averaged radial magnetic field of about 30--100 kG in the cores of some red giant stars from the observations of the rotational splitting asymmetry of their g-mode dominated mixed modes. 
For acoustic modes, the situation is more difficult because first-order perturbation is not adequate, even for slowly rotating stars. Perturbation theory to the second order has been applied to the p-mode splittings by \citet{sai81} ; \citet{sua06}; \citet{Dzi92}; \citet{Kje98}; and \citet{Chr00}. The second-order treatment has been compared with full 2-D calculations by \citet{ree06}, which shows nice agreement for slow/moderate rotation rates. And \citet{sou98} even extended the theory to the cubic order. The domain of validity of perturbative treatment of different orders can be found in \citet{ree22}. The implications for the rotational splittings of $\delta$\,Sct stars have been summarised in \citet{gou11}. \citet{giz16} studied the p-mode splitting asymmetry of the slowly rotating $\delta$\,Sct star KIC 11145123, for which the splitting asymmetry is attributed to stellar deformation caused by rotation. They inferred the flattening of the star to be $\Delta R/R \approx 10^{-6}$.

\citet{bed20} discovered a sample of high-frequency, young $\delta$\,Sct stars, for which some radial modes and $l=1$ modes can be identified from the echelle diagram: they form clear vertical ridges which curve to the right. Asteroseismic modeling is now possible for some $\delta$\,Sct stars \citep{mur23,scu23}. However, in most cases, the rotational splittings have not been modeled in detail, with the exception of a few works, e.g., \citep{pam03,sua05,bri07,zwi14}.



This article is organized as follows.
In Sections 2.1-2.3, we layout the theoretical expressions for the p-mode frequency corrections and asymmetry using the second-order perturbation theory, largely following the notations in \citet{sai81} and the development in \citet{Dzi92} and \citet{sou98} . We write the expressions explicitly so that readers can easily use them.
We then calculate the frequencies of MESA models representing $\delta$\,Sct stars in Sec 2.4.  We compare the theory with the observed splitting asymmetry from seven slowly rotating $\delta$\,Sct and $\beta$ Cep stars (Section 2.5). In Sec. 3, we discuss other factors and can affect the splitting symmetry, and then discuss future prospects (Sec. 4). We conclude with predictions on the rotational splitting asymmetry which can be tested by future observations (Sec. 5). 


\section{Second-order frequency corrections and splitting asymmetry of Acoustic Modes}

\subsection{Rotating stellar structure models from MESA}

We use MESA \citep{pax11,pax13,pax15,pax18} to build stellar structure models from the pre-main sequence. Stellar rotation is initiated at the zero-age main sequence (ZAMS). The settings for chemical mixtures \citep{gre93}, OPAL opacity tables \citep{Igl96}, metallicity ($Z=0.02$), mixing-length parameter ($\alpha_{\rm MLT}=1.8$) and exponential overshooting parameter $f_{ov}=0.02$ are the same as \citet{guo16,guo19}. Rotational mixings are turned off.

We consider uniform rotation $\Omega(r)=\Omega_{\rm rot}$, as asteroseismic observations suggest that the rotational profiles of intermediate and massive stars can be approximately described by solid-body rotation in the radiative envelope \citep{aer17}. The radial part of the centrifugal force, given by $-(2/3)\Omega_{\rm rot}^2 r$, is included in the gravitational acceleration to compute an effective gravitational acceleration $g_{\rm eff}$.

\subsection{Oscillation frequencies with second-order correction}

For slow or moderate rotation, we can use perturbation theory to study the effect of rotation on  stellar eigenfrequencies, with the ratio of rotational frequency to the non-rotational oscillation frequency being a small parameter $\mu=(\Omega_{\rm{rot}}/\omega_0)  \ll 1$. After taking into account the first and second-order rotational correction and the near-degeneracy effect, the oscillation frequencies in the observer's frame can be expressed as \citep{sua06}: 
$\omega=\omega_0+ (\omega_1 +\omega^{(1)}_{ab}) + (\omega_2 + \omega^{(2)}_{ab}) + O(\mu^3)$.
The frequency corrections $\omega_1, \omega_2$ are of the order $O(\mu)$ and $O(\mu^2)$, respectively. And the  $\omega^{1}_{ab}$ and $\omega^{2}_{ab}$ represent the near-degeneracy frequency corrections of corresponding orders (see the Appendix).

 The well-known first-order rotational correction $\omega_1$ can be expressed as:

\begin{equation}
\omega_1=m(1-C_{nl})\omega_0 \mu=m(1-C_{nl})\Omega_{\rm rot}, 
\end{equation}
where the Ledoux coefficient is $C_{nl}=\int^R_0{\rho r^2(2\xi_r\xi_h+\xi^2_h)dr}/I$, $I $ is the mode inertia: $I=\int{(\xi_r^2 +l(l+1)\xi_h^2)}\rho r^2 dr$. $\xi_r$ and $\xi_h$ are the usual  eigenfunctions for the radial and horizontal Lagrangian displacements.


To the second-order $O(\mu^2)$, the frequency correction is, 
\begin{equation}
\omega_2=(D_0+m^2D_1)\omega_0\mu^2.
\end{equation}

\noindent The second-order correction coefficients are $D_0=X_1+X_2+Z$,  
$D_1=Y_1+Y_2$ (Saio 1981).  The $Z$ term (due to the spherical part of the centrifugal force) has already been taken into account in the MESA evolution models (Sec. 2.1).  $X_2$ and $Y_2$ are due to the rotational non-spherical distortion, $X_1$ and $Y_1$ include the first-order perturbation of the displacement eigenfunction and the effect of inertia, and for p-modes $|X_2| \gg |X_1|$ and $|Y_2| \gg |Y_1|$. Thus we have, 
\begin{equation}
D_0\approx X_2, \ \  D_1\approx Y_2  \ \   \rm {(p\  modes)}.
\end{equation}
For p modes, the second-order frequency correction is then,
\begin{equation}
\omega_2 \approx (X_2+m^2Y_2)\Omega_{\rm rot}^2/\omega_0.
\end{equation}

Fig.\,3 shows the second-order rotational splitting coefficients $X_2$ and $Y_2$ as a function of stellar age. The calculation is based on the MESA structure model with $M=1.8\,{\rm M}_{\odot}, Z=0.02, V_{\rm ZAMS}=60$\,km~s$^{-1}$. It can be seen that these coefficients are nearly constant, and their values are very close to those from polytrope models with index $n=3$ and $\gamma=5/3$ \citep{sai81}. At later stages of the main sequence, the signatures of avoided crossings are evident: as the g-mode frequencies increase and approach those of the p-modes, this sequential approach from lower to higher-order p-modes is manifested in the sequential bumping of $Y_2$ and dipping of $X_2$. For these mixed modes, the polytrope values are no longer valid.

\label{sec:maths} 

For radial modes, the second-order frequency perturbation is \citep{Chr00}, 
\begin{equation}
\omega_2(l=0)=(-4/3)\Omega_{\rm{rot}}^2/\omega_0.
\end{equation}
\noindent Note that 
$|\omega_2(l=0)|$ is a decreasing function of frequency $\omega_0$. Thus high-order radial modes have smaller non-spherical-distortion frequency corrections. 

For non-radial modes,
\begin{equation}
\omega_2 \approx \frac{\Lambda-3m^2}{4\Lambda-3} J_c \frac{\Omega_{\rm{rot}}^2}{\omega_0},
\end{equation}
that is, $X_2=\frac{\Lambda}{4\Lambda-3} J_c$ and $Y_2=\frac{-3}{4\Lambda-3}J_c$, where $\Lambda=l(l+1)$, when comparing with equation (4). $J_c$ is an integral in \citet[Appendix B]{sou98}; see also \citet{sua06},
\begin{equation}
J_c = \left(\frac{1}{2I}\right)\int{\rho r^2 (d_1 F_1+d_2 F_2)},
\end{equation}
where $F_1$ and $F_2$ depend on mode eigenfunctions. And $d_1=r^2 u_2, d_2=r\frac{d(r^2 u_2)}{dr}$. In these expressions, $u_2$ depends on the non-spherical equilibrium structure $u_2=\phi_{22}/(\Omega_{\rm rot}^2 r^2)$. $\phi_{22}$ is the gravitational potential perturbation due to rotation, and it can be calculated by integrating the perturbed Poisson equation \citet[eq. 17]{sou98}, see also \citet{oua12}.

To be specific, the second-order corrections for $l=1$ and $l=2$ p modes are: 

\begin{equation}
\omega_2 (l=1) =\left\{
\begin{array}{rcl}
(2/5)\frac{\Omega_{\rm{rot}}^2}{\omega_0}  J_c & & {\textrm{if}\  m=0 }\\

-(1/5)\frac{\Omega_{\rm{rot}}^2}{\omega_0} J_c & & {\textrm{if}\  m=\pm 1 }
\end{array} \right.
\end{equation}

\begin{equation}
\omega_2 (l=2) =\left\{
\begin{array}{rcl}
(2/7)\frac{\Omega_{\rm{rot}}^2}{\omega_0} J_c & & {\textrm{if}\  m=0 }\\

(1/7)\frac{\Omega_{\rm{rot}}^2}{\omega_0} J_c & & {\textrm{if}\  m=\pm 1 } \\

-(2/7)\frac{\Omega_{\rm{rot}}^2}{\omega_0} J_c & & {\textrm{if}\  m=\pm 2 } \\
\end{array} \right.
\end{equation}
Note that $J_c$ is dimensionless, $m$-independent, and $l$-dependent.
For $l=1$ and $l=2$ modes, $J_c$ is an increasing function of radial order $n$. This increasing trend is partly cancelled by the term $\Omega^2_{\rm rot}/\omega_0$. But still, larger frequency corrections $\omega_2$ are expected for higher $n$ pressure modes.

\subsection{Theoretical pressure-mode asymmetry}

The $m^2$ dependence of the second-order frequency corrections induce an asymmetry in the dipole and quadrupole mode splittings. 
We consider $m=-1, 0, 1$ modes only. For convenience, we define the right and left frequency differences as: 
$\omega_R=\omega_{m=1} -\omega_{m=0}$ and $\omega_L=\omega_{m=0} -\omega_{m=-1}$.


Similar to \citet{deh17} and \citet{ong22},  We define a dimensionless asymmetry parameter
\begin{equation}
A=\frac{\omega_L - \omega_R}{\omega_L + \omega_R}=- \frac{ \omega_{+}+\omega_{-} -2\omega_0}{\omega_{+} - \omega_{-}}.
\end{equation}
Note that there is a sign difference in our definition, since p-mode rotational splittings tend to have $\omega_L > \omega_R$ (detailed below) but most observed magnetic-field-induced splittings tend to have $\omega_R > \omega_L$, if the magnetic axis is not too misaligned \citep{li22,mat23,das24}, and we find it convenient to work with positive numbers here.


Using equation (4), we can derive the asymmetry parameter for $m=1, 0, -1$ modes:
\begin{equation}
A_{m1,-1}= - \frac{\Omega_{\rm{rot}}}{\omega_0}\frac{ Y_2}{(1-C_{\rm{nl}})}
\end{equation}

For $m=2, 0, -2$ spittings, if we define frequency differences as
$\omega_R=\omega_{m=2} -\omega_{m=0}$ and $\omega_L=\omega_{m=0} -\omega_{m=-2}$, the corresponding asymmetry parameter $A_{m2,-2}$ is:
\begin{equation}
A_{m2,-2}= - 2\frac{\Omega_{\rm{rot}}}{\omega_0}\frac{ Y_2}{(1-C_{\rm{nl}})}
\end{equation}
Note that since $Y_2$ is strictly negative, both asymmetry parameters $A_{m1,-1}$ and $A_{m2,-2}$ are positive, according to second-order frequency correction expressions. Note that eqn. (12) and (11) do not indicate that $A^{l=2}_{m2,-2}$ is twice that of $A^{l=1}_{m1,-1}$, since both $Y_2$ and $C_{nl}$ are $l$-dependent. 

Thus, the most prominent feature of the p-mode rotational splittings is positive asymmetry ($A>0$), i.e., the left frequency spacing $\omega_{\rm L}$ is always larger than the right spacing $\omega_{\rm R}$.

\subsection{Calculated evolution of oscillation frequencies, rotational splitting coefficients and asymmetry parameter A}

In  Fig.\,1, we show the evolution of oscillation frequencies of $l=0,1$ (upper panel) and $l=0,2$ (lower panel) acoustic modes as a function of stellar age (in units of $10^8$\,yr), effective temperature ($T_{\rm eff}$), and stellar radius (R). It is based on a stellar structure model with $M=1.8$\,M$_\odot$, $Z=0.02$, $V_{\rm ZAMS}=60$\,km~s$^{-1}$. The $l=1$ splitting triplets $m=-1,0,1$ and $l=2$ quintuplets $m=-2,-1,0,1, 2$ are obviously asymmetric. It can be observed that the asymmetry $A_{m1,-1}$ is positive, and higher radial-order modes have stronger asymmetry.

\begin{figure*}
	\includegraphics[angle =90,width=1.9\columnwidth]{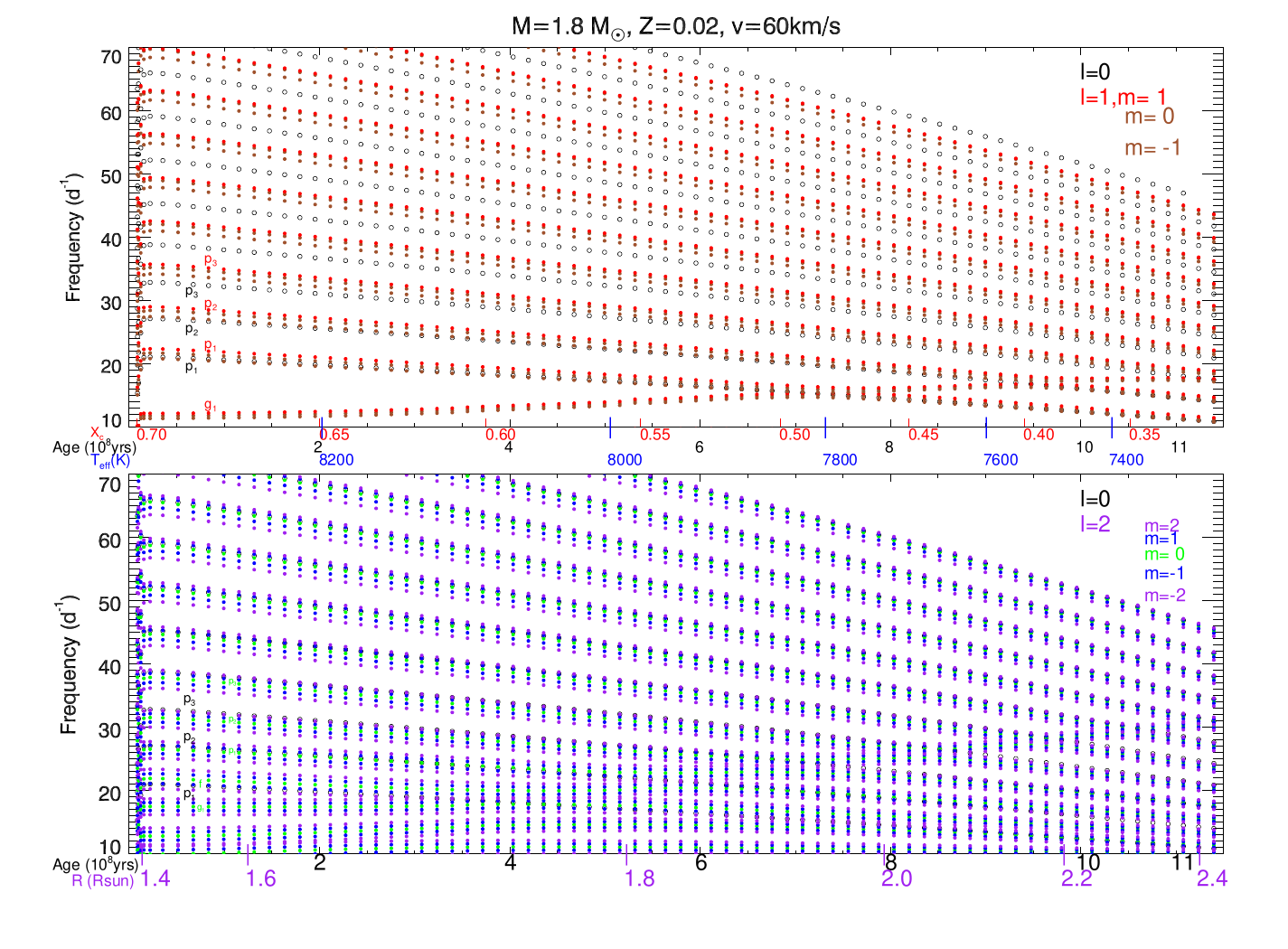}
    \caption{Evolution of oscillation frequencies with asymmetric rotational splittings for a typical 1.8\,$\mathrm{M_{\odot}}$ $\delta$\,Sct star with solar metallicity. The $x$-axes include stellar age, effective temperature ($T_{\mathrm{eff}}$), and stellar radius (R).}
    \label{fig:6}
\end{figure*}


\begin{figure*}
	\includegraphics[angle =0,width=1.9\columnwidth]{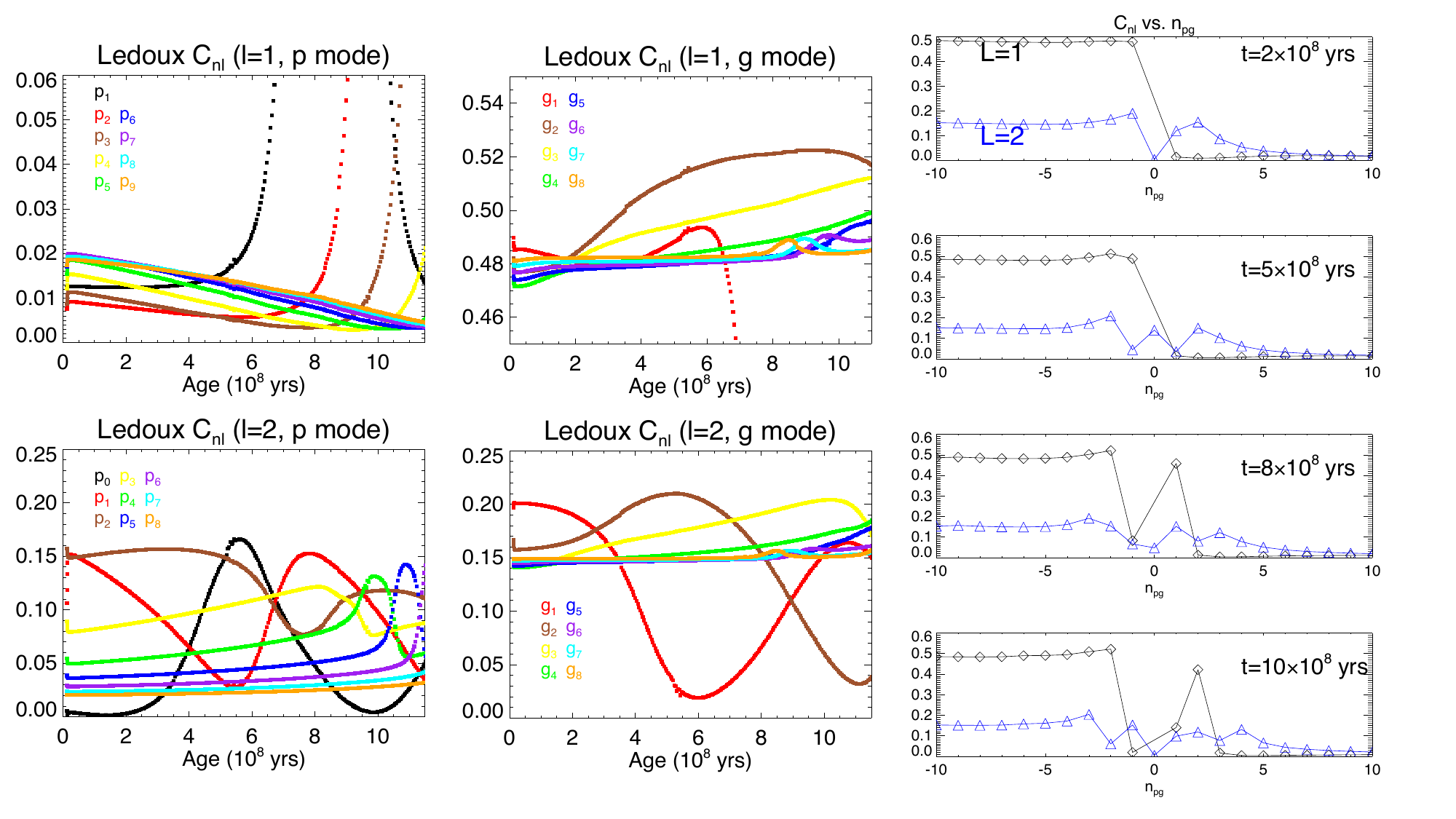}
    \caption{Ledoux coefficients $C_{\rm nl}$ of $l=1$ and $l=2$ modes for a stellar model with $M=1.8$\,M$_\odot$, $Z=0.02$. Left panels show the variations with stellar age, and right panels show the variations with radial order $n$ at different time snap shots (using the GYRE notation of \citealt{tow13}), $n_{\rm pg} <0$ for g modes and $>0$ for p modes).}
    \label{fig:example_figure5}
\end{figure*}


\begin{figure}
	\includegraphics[angle=90,width=1.0\columnwidth]{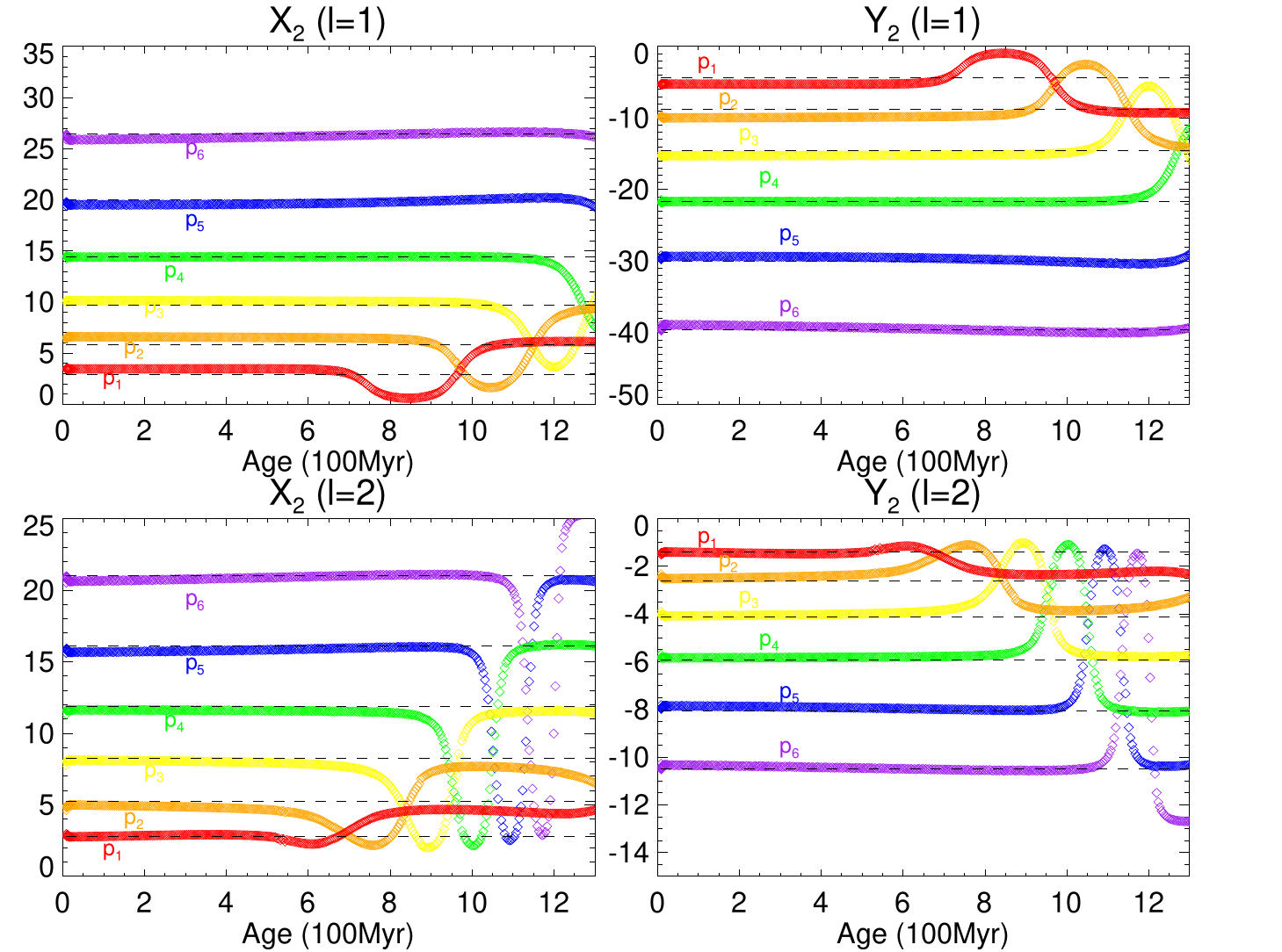}
    \caption{Second-order rotational splitting coefficients $X_2$ and $Y_2$ calculated from of a $M=1.8$\,M$_\odot$ MESA model, with initial rotational velocity $v_{\rm eq} = 60$\,km~s$^{-1}$ at the ZAMS. The horizontal dashed lines are results based on polytrope models from \citet{sai81}.}
    \label{fig:example_figure4}
\end{figure}

\begin{figure*}
	\includegraphics[angle =0,width=1.9\columnwidth]{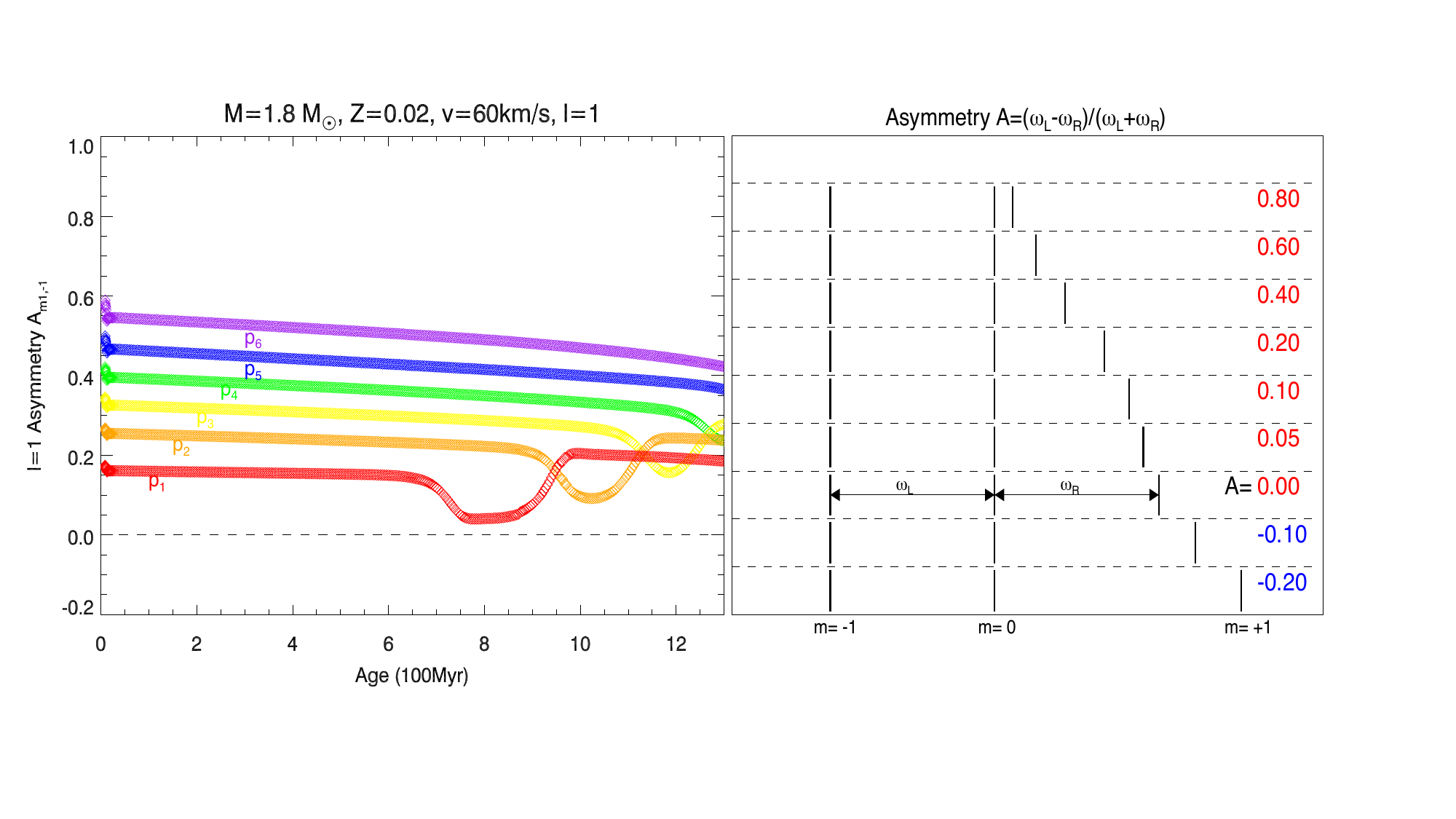}
    \caption{ \textbf{Left: }The variation of the splitting asymmetry parameter of dipole modes $A_{m1,-1}$ for a $M=1.8$\,M$_\odot$, $Z=0.02$ MESA model. \textbf{Right}: A series of asymmetric splitting triplets with decreasing asymmetry parameter $A$.}
    \label{fig:exfigure0}
\end{figure*}

In Fig.\,2, we show the first-order rotational splitting coefficients (Ledoux constant) for both dipole and quadrupole modes. 
For dipole modes, the $C_{n,l=1}$ of pressure modes are always very small at the ZAMS, being less than $2$ per~cent. As the star evolves, their values are slowly decreasing, until an avoided crossing occurs (sequentially to $p1$, $p2$, ...), and the values start to become much larger, i.e., evolve towards the asymptotic g-mode's value $1/\Lambda$. 

For gravity modes, the initial $C_{nl}$ start with an value around $0.48$, close to the asymptotic value $1/l(l +1)=0.5$. These remain almost constant, until avoided crossing occurs and the values decrease sharply. This can also be seen in the right-most panel which shows $n_{\rm pg}$ (radial order) vs $C_{nl}$, where the $C_{nl}$ of the g modes have very small variations except when avoided crossings occur sequentially to $g_1, g_2, ...$. For instance, at $t=8 \times 10^8$ yr, the $C_{nl}$ of the $g_1 (n_{\rm pg}=-1)$ mode drops from 0.5 to 0.1 due to coupling to p modes.
 Low radial-order g modes have a larger $C_{nl}$ variation, and higher-order g modes are much less affected. The Appendix of \citet{kur14} derived the asymptotic estimate of $C_{nl}$ for dipole g modes based on the theory of \citet{tak05}.

For $l=2$ modes, the $C_{nl}$ of g modes are similar although the asymptotic value is then $1/l(l+1)=1/6 \approx 0.167$.
For pressure modes, however, the Ledoux coefficients are not so small. The f mode starts with a Ledoux value $\approx 0$ but its value reaches $\approx 0.167$ at around 600 Myrs. The $C_{nl}$ of $p_1$ and $p_2$ start with a large value of 0.15 at the ZAMS, but then decrease.
Note that the $l=2$ mode avoided crossing happens earlier than $l=1$; we thus see p modes possessing larger $C_{nl}$ at the early phase of the main sequence.  For $p_1$ modes, $C_{nl}$ reaches 0.15 at $800$ Myr. After that, $p_3$ $p_4$, and $p_5$ sequentially reach their largest values as they couple to g modes.
For both $l=1$ and $l=2$, the $C_{nl}$ of g modes are close to the asymptotic values of $1/2$ and $1/6$ for high-order modes $n_{\rm pg} <-5 $.

In Fig.\,3, we show the second-order rotational splitting coefficients $D_0 \approx X_2$ and $D_1 \approx Y_2$  of $l=1, 2$ pressure modes for the same MESA stellar model. Starting from the ZAMS, these coefficients have near-zero variations, the values almost follow the constants tabulated in \citet{sai81} for an $n=3$ polytrope (dashed lines). $X_2$ are strictly positive, and $Y_2$ are strictly negative. Also, higher-order p modes have larger $X_2$ and $|Y_2|$. At later stages of main sequence, we can see the signatures of avoided crossings:  the increasing g-mode frequencies are getting close to the p modes, and this happens sequentially from low to higher-order p modes, manifested as the sequential bumping in $Y_2$ and dipping in $X_2$. For these mixed modes, the polytrope values are no longer valid.

The sign of $X_2$ and $Y_2$ have implications on the splitting asymmetry. First, $Y_2$ is negative, thus both prograde and retrograde modes have negative second-order frequency shifts. Since the $1^{\rm st}$-order Ledoux splitting is symmetric, we have $\omega_L > \omega_R$ for second-order splittings, i.e., positive asymmetry parameters $A_{m1,-1}$ and $A_{m2,-2}$.


In Fig.\,4, we present the dimensionless asymmetry parameter of dipole modes $A_{m1,-1}$ and its evolution with stellar age. It can be seen that the asymmetry $A_{m1,-1}$ is slowly evolving with stellar age, but sensitive to radial order $n$. When avoided crossing kicks in, the asymmetry decreases significantly. This happens to the fundamental radial mode ($p_1$) at $7\times 10^8-10\times 10^8$ yrs, and then around $9\times 10^8 -11.5\times 10^8$ yrs to first-overtone radial mode ($p_2$).

The right panel helps readers to compare the $A$ values with the actual appearance of the $l=1$ triplets.

\subsection{Observational evidence of the p-mode splitting asymmetry}

Thanks to space telescopes, we have accumulated a significant number of observations of p-mode splittings in $\delta$\,Sct stars, with very high frequency resolution.

We have collected data on 
three $\delta$\,Sct
stars and four 
$\beta$\,Cep stars from the literature, all of which exhibit well-measured p-mode rotational splittings, and are slowly rotating ($v\sin i \lesssim 30$ km~s$^{-1}$). Detailed information on the stellar parameters and the rotational splittings can be found in Table A1. The slow rotation of these stars places them within the valid regime for applying second-order perturbation to rotation. As previously noted, the dominant effect of non-spherical distortion governs the p-mode splittings, allowing for the measurement and model comparison of the induced asymmetry. The high frequency resolution ensures that the splitting asymmetry parameter is determined with exceptional precision, even for these slowly rotating stars.

In Fig.\,5 we display all the measurements of p-mode rotational-splitting asymmetries in the seven stars. It is immediately apparent that almost all the splittings have positive asymmetry $A$. The symbols predominantly occupy the upper half of the plane, with only one mode as a significant exception. This consistent positivity is a direct consequence of the rotational non-spherical distortion.

In Fig.\,6 we quantitatively illustrate the asymmetry parameter $A$, calculated from models ($l=1$ in black and $l=2$ in gray). The model curves extend from right to the left, reflecting the evolution of the star. Different curves are for different radial orders and we plot all the models within the $2\sigma$ error boxes of mass and radius. The observed splitting asymmetries in KIC~10080943 (three blue symbols) likely correspond to $l=1$, allowing for direct comparison with the models (black lines). The stellar parameters of the $\delta$\,Sct star in KIC100809043 are constrained to $M_a=2.0(0.1), R_a=2.9(0.1)$ or $M_b=1.9(0.1), R_b=2.1(0.2)$ in solar units, with rotational velocity $v_{\rm eq} =14.4$\,km~s$^{-1}$ \citep{sch15}.  Rather than seeking a single optimal model, we present various potential models for a schematic quantitative comparison. While there is general agreement, we find that the model asymmetry from our calculations seems to be larger than the observations. Additional factors reconciling this discrepancy are discussed in the next section.
\begin{figure}
	\includegraphics[angle=0,width=1.0\columnwidth]{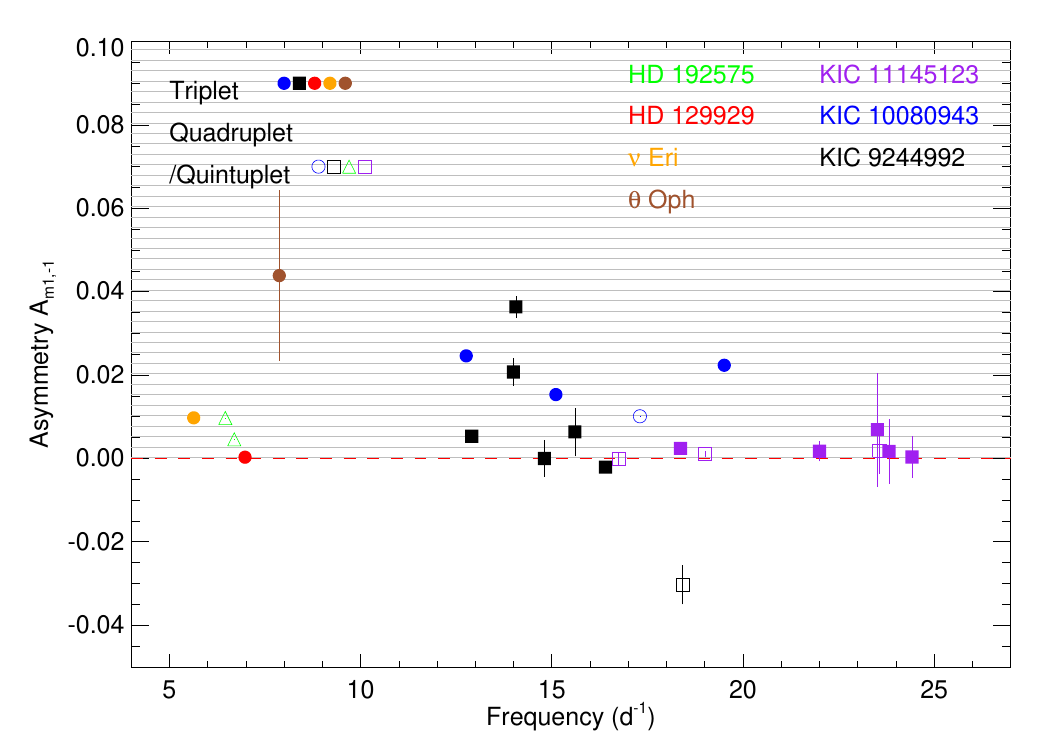}
    \caption{The observed asymmetry parameter A for $m=1,0,-1$ modes from six stars is colour-coded. Filled symbols represent triplets, whereas open symbols indicate quadruplets or quintuplets. The majority of these measurements are situated in the upper, positive half of the plane.}
    \label{fig:examplef1}
\end{figure}


\begin{figure}
	\includegraphics[angle=90,width=0.9\columnwidth]{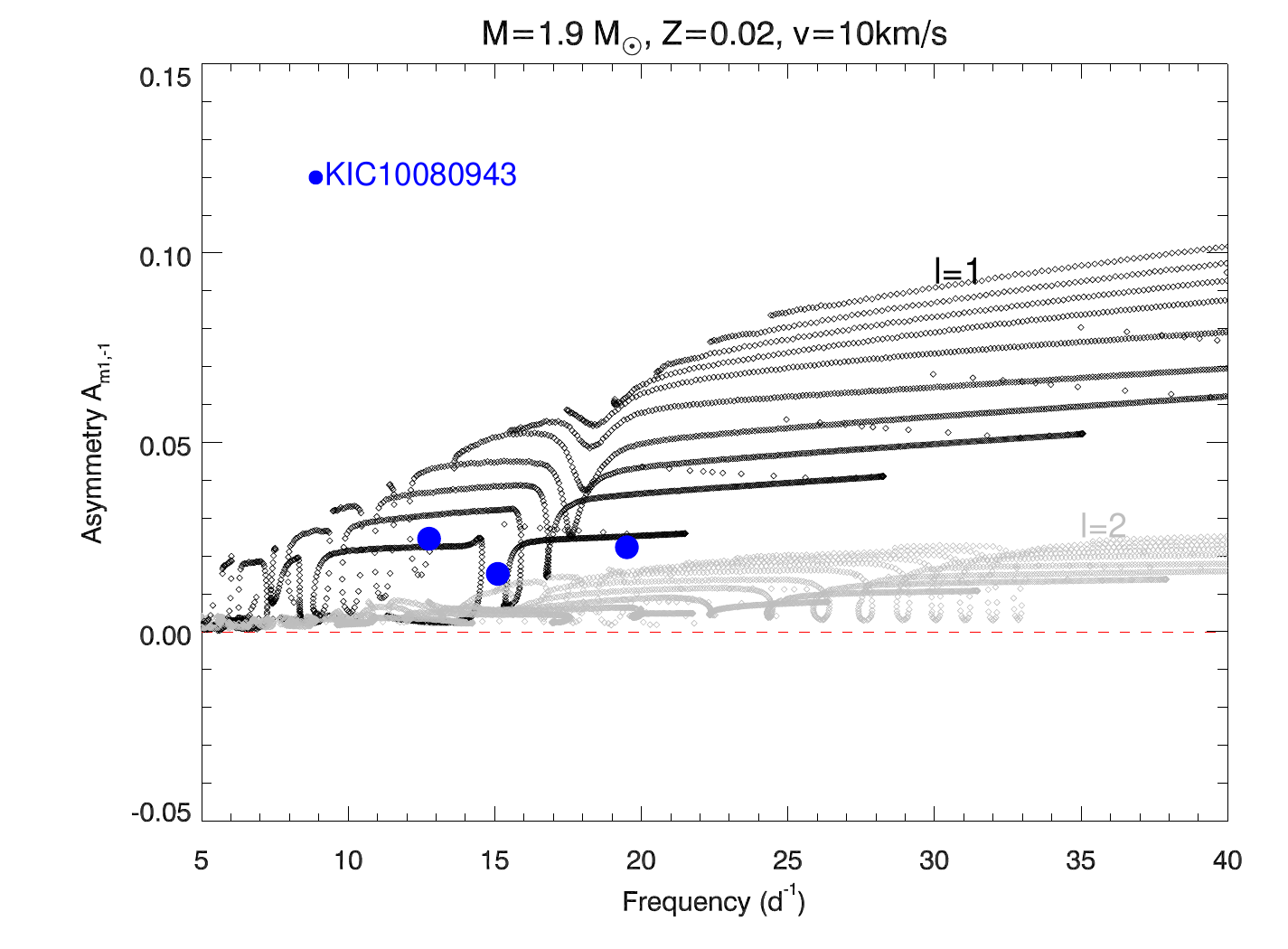}
    \caption{The observed asymmetries in the rotational splitting triplets of KIC~10080943 (blue symbols) are superimposed on the theoretical asymmetry parameters for $l=1$ modes (in black). Each line corresponds to a mode of a different radial order. Models covering a range of stellar ages are displayed, and the time evolves from right to the left. For comparison, the theoretical $l=2$ mode asymmetries are also depicted (in grey). }
    \label{fig:exame_figure2}
\end{figure}

\begin{figure}
	\includegraphics[angle=0,width=1.2\columnwidth]{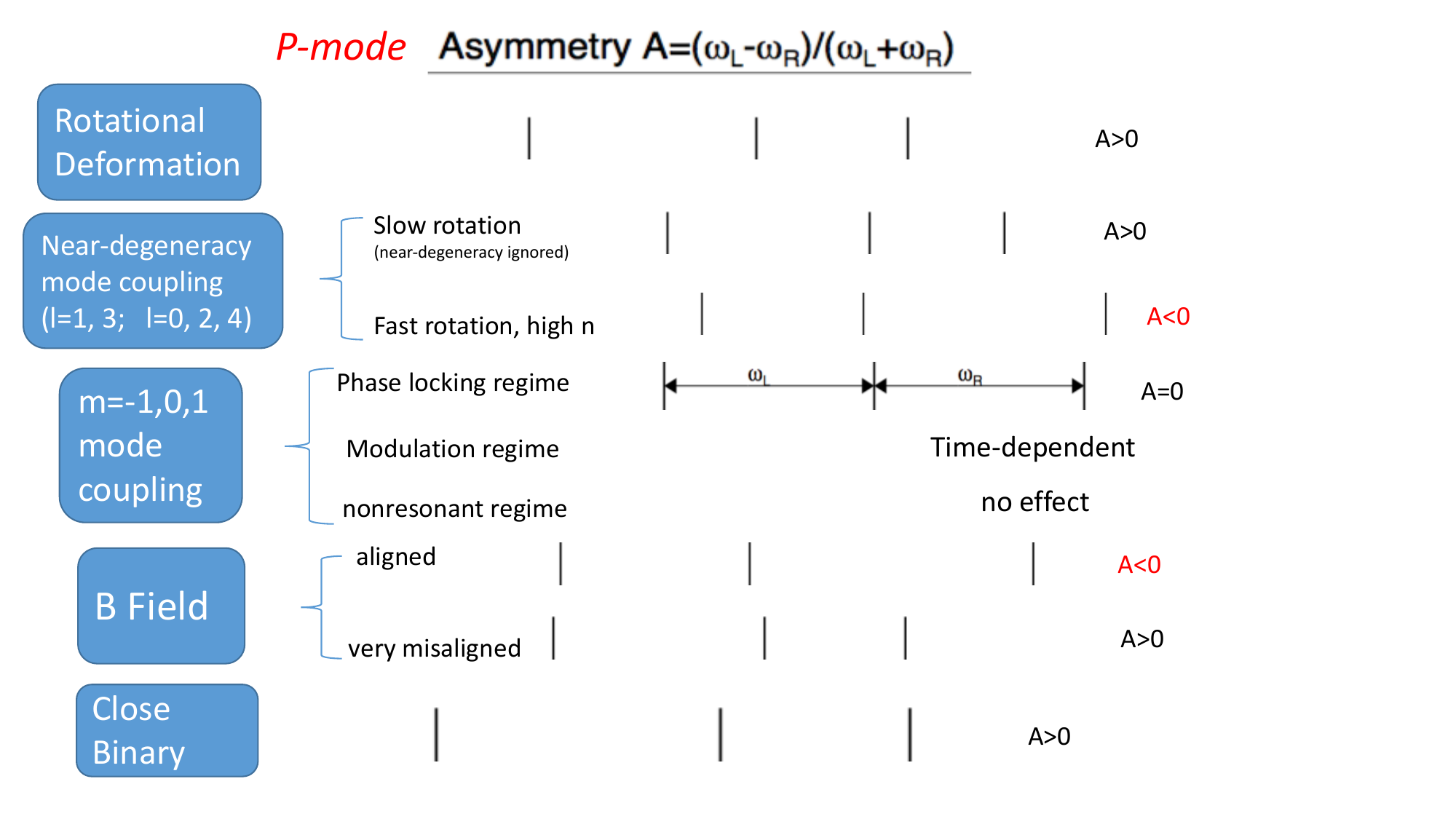}
    \caption{Summary of factors affecting the p-mode asymmetry parameter A. The list includes rotational deformation, magnetic field (denoted as B fields), resonant mode coupling among $m=1,0,-1$, near-degenerate mode coupling, and tidal effect in close binaries.}
    \label{fig:example_figure}
\end{figure}

Note that, observationally, we usually look for almost equally spaced multiplets. When these splittings align with theoretical predictions based on rotation, they can be used to determine the rotational velocity. Consequently, there is a pronounced bias towards identifying rotational multiplets for slowly rotating stars. Indeed, all the p modes in Fig.\,5 have relatively low frequencies, indicating a slightly evolved, slower rotation phase of the star. Notably absent in the literature is the identification of non-equally-spaced rotational splittings among young stars and higher frequency p-modes.


\section{Other factors affecting the splitting asymmetry}

In Fig.\,7, we show a schematic summary of the factors that can affect the observed p-mode asymmetry. Except for the rotational deformation, other factors include a magnetic field, resonance mode coupling, tidal deformation in a close binary and the near-degeneracy effect. We discuss each of these factors below.

\subsection{Magnetic effect on the splitting asymmetry}

Most $\delta$\,Sct stars do not exhibit detectable magnetic fields at their surfaces. The first evidence of a main-sequence magnetized $\delta$\,Sct star was reported by \citet{nei15} in HD~188774, where a magnetic field of $<100$\,G was inferred from spectro-polarimetric measurements. \citet{tho23}
found no magnetic field signatures in a sample of $\delta$\,Sct/$\gamma$\,Dor hybrid pulsators.

However, strong magnetic fields may still exist near the core.
Observations of g-mode dominated dipole mixed modes in red giants mostly show negative asymmetry, with $\omega_L < \omega_R$, i.e., $A<0$ (or $a>0$ as in \citealt{li22}; \citealt{deh23}). A dipole field has to be highly inclined (with an inclination angle $\beta > 54.7^\circ$) for the splitting to exhibit positive asymmetry ($A>0$), as shown by \citet{mat23} and \citet{das24}. This prominent feature of negative asymmetry has been employed to infer an averaged strength of $30-100$\,kG for the radial components of the magnetic field.

For p modes or p-mode dominated mixed modes in main-sequence stars, the indirect effects of a magnetic field are anticipated to be larger than the direct effects \citep{Mat21}. The influence of Lorentz force on the stellar equilibrium structure, and consequently on the mode propagation and eigenfrequencies needs to be considered \citep{gou90}. The relative frequency shift of p modes scales as $(v_A/c)^2$ (the ratio of Alfven velocity to sound speed). Since p mode frequency is substantially higher than the Alfven frequency, this effect is considerably smaller than that caused by rotational non-spherical distortion. 
We conduct an order of magnitude estimation of the magnetic frequency shift for the $l=1$ splittings, following \citet{Mat21}.
Assuming the magnetic field axis is aligned with the pulsation axis,

\begin{equation}
\delta\omega_{mag} \approx \frac{D_{lm}}{2}\frac{\int^R_0{(v_A/c)^2 dr/c}}{\int^R_0{dr/c}},
\end{equation}
where $D_{lm}=\int_0^{\pi}\sin^2\theta |Y^m_l|^2\sin \theta d\theta$ and the Alfven velocity $v_A=B^i/\sqrt{2\pi\rho}$ and ($i=\theta, \phi$). By using a typical mean value of $B^i \approx 10^5$\,G \citep{bug21}, which aligns with the measured strength in the core of red giants and near the core of the B star HD 43317 \citep{lec22} under the assumption of magnetic flux conservation. We find the frequency shift for a 1.9\,M$_\odot$, $R=2.1$\,R$_\odot$ model with oscillation frequency $f=15$\,d$^{-1}$ (KIC\,10080943) to be $\delta\omega/\omega \approx 10^{-7}-10^{-6}$, which translates to an asymmetry of $A \approx -10^{-4}$. Although this magnetic field-induced frequency shift slightly reduces the asymmetry, it reconciles, but does not fully resolve, the discrepancy illustrated in Fig.\,6.

\subsection{Binary effect on the asymmetry $A$}

We expect a larger splitting asymmetry in close binaries due to tidal distortion, in addition to the rotational deformation. Following \citet{sai81}, we can estimate the effect of tidal distortion on the oscillation frequencies for very close binaries with synchronised orbits. The second-order frequency correction term is now $s(X_2+m^2Y_2)$, where $s=(1+\frac{3}{2}\frac{q}{1+q})$, $q=M_2/M_1$, with $s=1$ being the single star scenario.
Thus, the corresponding theoretical splitting asymmetry contains an additional factor $s$: 
\begin{equation}
A=-s \left(\frac{\Omega_{\rm{rot}}}{\omega_0}\right)\frac{ Y_2}{(1-C_{\rm{nl}})}.
\end{equation}

Since $s>1$, we expect that the acoustic mode asymmetry $A$ becomes larger in close binary systems with significant tidal distortion. However, none of our stars are in very close binaries with synchronised rotation, making this theory not directly applicable to the stars studied in this paper.  KIC~10080943 is a binary, and it has an eccentric, 15-d orbit. $\theta$ Oph is in a triple system, with the inner binary in an eccentric, 56.7-d orbit. The effect of tidal distortion may be significant near periastron passage. A detailed calculation is beyond the scope of this paper.

\subsection{Resonant mode coupling among $m=-1, 0, 1$ triplets }

Mode coupling among the modes in the triplets satisfies the four-mode coupling resonance condition 
$2\omega_{m=0} \approx \omega_{m=1} + \omega_{m=-1}$\footnote{This is a cubic-order four-mode coupling with two same $m=0$ modes.} . 
We denote the frequency detuning from this exact resonance by $\delta \omega$.
The behaviour of coupled-mode system can be described by the amplitude equation framework with third-order nonlinearity \citep{gou94,buc95,now01}.

Stationary solutions and instabilities of this $1:1:1$ resonance system have been extensively studied \citep{buc95}. 
According to \citet{gou98}, there are three distinct regimes, depending on the frequency detuning $\delta \omega$ and linear growth/damping rate of the $m=0$ mode $\gamma_0$. Firstly, the system can reach a stationary state, with constant amplitude and phase (frequency) if $\delta\omega \lesssim \gamma_0$. This frequency/phase locking regime will induce equally-spaced splittings ($\omega_L= \omega_R$), with asymmetry $A=0$.
The second regime involves modulation if $\delta\omega \gtrsim \gamma_0$, corresponding to periodic solutions. The predicted splittings are thus periodic. In the third regime, if the detuning is much larger than the growth rate, the system is essentially non-resonant, and modes can be treated independently.

It is likely that the six $\delta$\,Sct systems we studied mostly fall into the non-resonant or modulation regimes, since it requires very fine frequency detuning $\delta\omega$ for the frequency-locking regime condition to be satisfied.
It would be enlightening to study the amplitude/phase variations of these $l=1$ triplet and $l=2$ multiplets observationally. Observed amplitude and phase (frequency) modulation of $m=1,0,1$ modes in sub-dwarf B stars and white dwarfs have been reported by \citet{zon16}, which corresponds to the modulation regime.

Finally, if the pulsation axis is misaligned with the rotation axis, the observed oscillation mode in the Fourier spectrum will exhibit side peaks separated by the rotational frequency. This applies to the oblique pulsator roAp stars \citep{kur82}, as well as the tidally perturbed close binary systems \citep{han20,ful20,van23}. In these cases, the asymmetry parameter $A$ is zero.

\subsection{Near-degeneracy mode coupling effect}

This effect occurs between modes with spherical degree $l$ and $l+2$ with the same $m$ (extending to $l+4$ if the cubic order is considered). It becomes stronger for higher radial orders and for faster rotation.
The appearance is that if the two modes are too close, they begin to repel each other. The p-mode positive asymmetry $A$ can be reduced or even inverted to negative values (e.g. Fig.\,A1). This coupling can strongly alter the radial mode and $l=2$ mode frequencies of higher order. This phenomenon requires a technical description therefore we detail the rotational near-degeneracy mode coupling effect and how it affects the mode frequency, large frequency separation and splitting asymmetry in the Appendix.

\section{Discussion and Future Prospects}

$\delta$ Scuti stars are typically fast rotating, and thus, the effects of rotational non-spherical distortion are crucial in the analysis of their oscillation frequencies. Traditionally, nearly equally spaced peaks in Fourier spectra are identified as possible rotational splittings. Leveraging second-order perturbation theory, our research aims to aid observers in detecting and characterizing splittings in the data, extending beyond nearly equally spaced peaks to include asymmetric ones as well.

A large sample of $\delta$\,Sct and $\beta$\,Cep stars with precisely measured oscillation frequencies and amplitudes from {\it TESS} has become available \citep{hey24}. Rotational splittings can be identified and the asymmetry can be extracted. It is desirable to verify the correlations predicted in this work. It is also important to study the mode excitation preferences, mode energy distribution in the rotational splittings \citep{lee95}. On the theoretical side, investigating subtle cubic-order effects and both radial and latitudinal differential rotation remains valuable \citep{oua12a,hat19}.

There are many $\delta$\,Sct or $\beta$\,Cep stars in eclipsing binaries \citep{gau19}. 
The p-mode asymmetry could potentially be used to infer the rotational and tidal distortion of the star. Some stars have already demonstrated evidence of rotational splittings \citep{sou23}. However, it may not be easy to find many systems in which p-mode splitting asymmetry can be measured, since the $m=0$ modes in the edge-on systems have lower visibility due to geometric cancellation. In some tidally perturbed binaries, the pulsation axis does not align with the rotation axis. Orbital modulation of oscillation amplitudes creates side peaks that complicate the measurement of true rotational splitting asymmetry.

For fast rotators, perturbative treatment becomes inadequate, necessitating full two-dimensional calculations. Two-dimensional stellar structure models, such as those computed with ESTER \citep{rie16}, and oscillation codes like TOP \citep{ree06} and ACOR \citep{oua12a}, have been employed to model a select few $\delta$\,Sct stars \citep{bou20,ree21}. This approach is now being extended to $\beta$\,Cep stars \citep{mom23}. 

In a future publication, we plan to present a finer grid of oscillation frequencies with second-order perturbation and near-degeneracy effect. This grid will be applied to the seismic modeling of young $\delta$\,Sct stars characterized by very regular patterns and clearly identified modes.
With continuous improvements in both observational capabilities and theoretical models, the field of p-mode asteroseismology for fast-rotating stars is poised for significant advancements.
\section{Conclusions}

We examined the rotational splitting asymmetry of pressure modes in $\delta$\,Sct and $\beta$\,Cep stars, detailing the theoretical expressions for the second-order perturbative effects on the splittings (Sec. 2), comparing these with observations (Figs 5 and 6), and considering other influencing factors (Fig.\,7).

We summarise our results as follows: 

\textbf{(1)} The pressure-mode frequency shift due to rotation is dominated by the non-spherical distortion, which produces positive asymmetry $A>0$ for $m=-1,0,1$ modes (Figs 1 and 4). This positive asymmetry tends to be more pronounced for higher order modes. 

\textbf{(2)} The second important factor affecting rotational splitting asymmetry is rotational near-degeneracy mode coupling between modes with $l$ differing by two. This effect is more significant for faster rotating stars. The result is to reduce the positiveness of asymmetry $A$, and even make $A$ negative for higher radial order $n$ modes (Fig.\,A1).

\textbf{(3)} Our collected $\delta$\,Sct and $\beta$\,Cep stars are slowly rotating with relatively low radial order p modes. Thus we expect the dominant factor is the non-spherical rotational distortion which induces positive asymmetry, in agreement with observations (Fig.\,5).

\textbf{(4)} A wealth of stars showing rotational splittings is present in the {\it Kepler} and {\it TESS} data sets. We predict the following trends in the splittings asymmetry. There would be a correlation between the splitting asymmetry $A$'s positiveness/negativeness and the splitting magnitude (proportional to rotation rate). Slow rotators and lower-frequency p modes typically show small positive asymmetry, with this positiveness increasing with oscillation frequency (Figs 4 and 6). Very low frequency modes will show $A$ much closer to zero. Conversely, faster rotating stars and higher frequency p modes will show smaller positive $A$ and even negative asymmetries. It will be interesting to check this correlation with more data.

\textbf{(5)} We expect the magnetic field will generally reduce the positiveness of asymmetry parameter, if the field is not very misaligned. But it is not the most important factor, given a typical field strength of $10^4-10^5$\,G near the convective-core boundary, and the induced frequency shift is at least one order of magnitude smaller than the aforementioned second-order effect.  However, if the non-spherical distortion and near-degeneracy effects are accurately modeled and removed, it may become feasible to detect magnetic splittings in the residual signal, which should scale with the p-mode frequency as 
$\propto \omega^1$ \citep{bug21}.

\textbf{(6)} Resonant mode coupling among the $m=-1, 0, 1$ modes could be significant if the frequency detuning is small enough. We may expect periodic modulation in the splitting frequencies, and even frequency-locking (asymmetry $A=0$). Additional stellar deformation from the tidal potential of the companion star is expected to increase the p-mode splitting asymmetry. It would be desirable to examine these splittings in an ensemble of close binaries with p-mode pulsators and compare them to those of singular-star pulsators.

\section*{Data Availability}
 The data generated in this research will be shared on reasonable request to the corresponding author.
 
\section*{Acknowledgements}
We thank Gerald Handler, Lisa Bugnet, Vincent Vanlaer, Dario Fritzewski and Lucas Barrault for helpful discussions.The research leading to these results has received funding from the European Research Council (ERC) under the Horizon Europe programme (Synergy Grant agreement N$^\circ$101071505: 4D-STAR).  While funded by the European Union, views and opinions expressed are however those
of the author(s) only and do not necessarily reflect those of the European Union or the European Research Council. Neither the European Union nor the granting authority can be held responsible for them. T.R.B. acknowledges support from the Australian Research Council through Laureate Fellowship FL220100117.









\appendix

\section{Near-degeneracy effect}

For a rotating star with spin frequency $\Omega_{\rm rot}$, when two oscillation modes are close to each other $\omega_a \approx \omega_b$, $| \omega_a - \omega_b| \le \Omega_{\rm{rot}}$, and if they also satisfy the selection rules $l_a=l_b$ or $l_a=l_b\pm 2$, and $m_a=m_b$ \citep{sua06}, the rotational near-degeneracy mode coupling effect becomes significant. The mode eigenfunctions are represented by a sum of both modes, and new eigenfrequencies are derived from a quadratic equation. Visually, this appears as the two modes beginning to repel each other. 
\citet{Das02} explored the impact of near-degeneracy on amplitude ratios and phase differences in multi-colour photometry.
\citet{sua07} analyzed how this effect modifies the Petersen diagram (Oscillation Period ratio vs. Period) of Cepheids and later applied it to $\beta$\,Cep stars \citet{sua10}.
Here we focus on how this effect alters the splitting asymmetry $A$ and the large frequency separation.

We calculate the oscillation frequencies at different levels of sophistication: 

$\omega_0$

$\omega_{\rm lin}=\omega_0+\omega_1$

$\omega_{\rm 2nd}=\omega_0+\omega_1+\omega_2$

$\omega_{\rm 2ndcoup}=\omega_0+\omega_1+\omega_2+\omega^{(2)}_{ab}$,

 which corresponds to the non-rotational frequency, the linear and second-order corrected frequencies, and the second-order corrected frequency with near-degeneracy mode coupling, respectively. The meaning of these frequencies follows the same notation in Section 2.

Figure 6 of \citet{zwi14} presents a `fork' plot of oscillation frequencies for a $M=1.7$\,M$_\odot$ model across these levels of complexity (see also \citealt{pam03}). They label the $\omega_{\rm 2nd}$ as `non-spherical distortion' and $\omega_{\rm 2ndcoup}$ as `three-mode coupling'\footnote{They denote the level $\omega_{\rm 2ndcoup}$ as `three-mode coupling' since they considered $l=0, 2, 4$ coupling. But we find the result is essentially the same if only consider $l=0$ and $2$.}.
For $\delta$\,Sct stars, if $\omega_{n,0}$ denotes the radial mode ($n=0$) frequency with radial order n, we find a mode frequency proximity relation $\omega_{n,0} \approx \omega_{n-1,2}$. The difference is termed the `small frequency separation', and denoted by $\delta_{02}$. 
For high-order p modes, these mode pairs draw even closer, as implied by the asymptotic relation $\omega_{n, l+2} \approx \omega_{n-1,l}$. For $l=1$ modes, the closest $l=3$ modes are further away compared to the $l=0, 2$ case, thus $\delta_{13} > \delta_{02}$.

Note that we consider the second-order near-degeneracy frequency correction $\omega^{(2)}_{ab}$\footnote{Note that $\omega^{(1)}_{ab}=0 $ in the case of $l_a = l_b\pm 2$. Please refer to \citet{deh17}  for the first-order near-degeneracy effect. $\omega^{(1)}_{ab} $, depends on eigenfunctions of mode a and b, the smaller frequency difference, larger near-degenerate correction. \citet{deh17} studied the first-order near-degeneracy effect for mixed modes of solar-like oscillating stars. The first-order effect is important in the study of mixed mode frequencies.} to these mode pairs, both with azimuthal $m=0$.

Focusing on the $l=1$ and $l=2$ modes shown in Figure 6 of \citet{zwi14}, linear frequency corrections strictly result in symmetric splittings ($A=0$). Second-order corrections generate positive asymmetric splittings, with $\omega_L$ larger than $\omega_R$. With the near-degeneracy effect included, the $l=2, m=0$ modes begin to repel the closest $l=0$ modes, shifting the $l=0$ modes to higher frequency and the $l=2,m=0$ modes to lower frequency, thus reducing the splitting asymmetry $A_{m1,-1}$.
The splitting asymmetry is strongly affected for higher-n modes.

\begin{figure*}
	\includegraphics[angle=90,width=2.2\columnwidth]{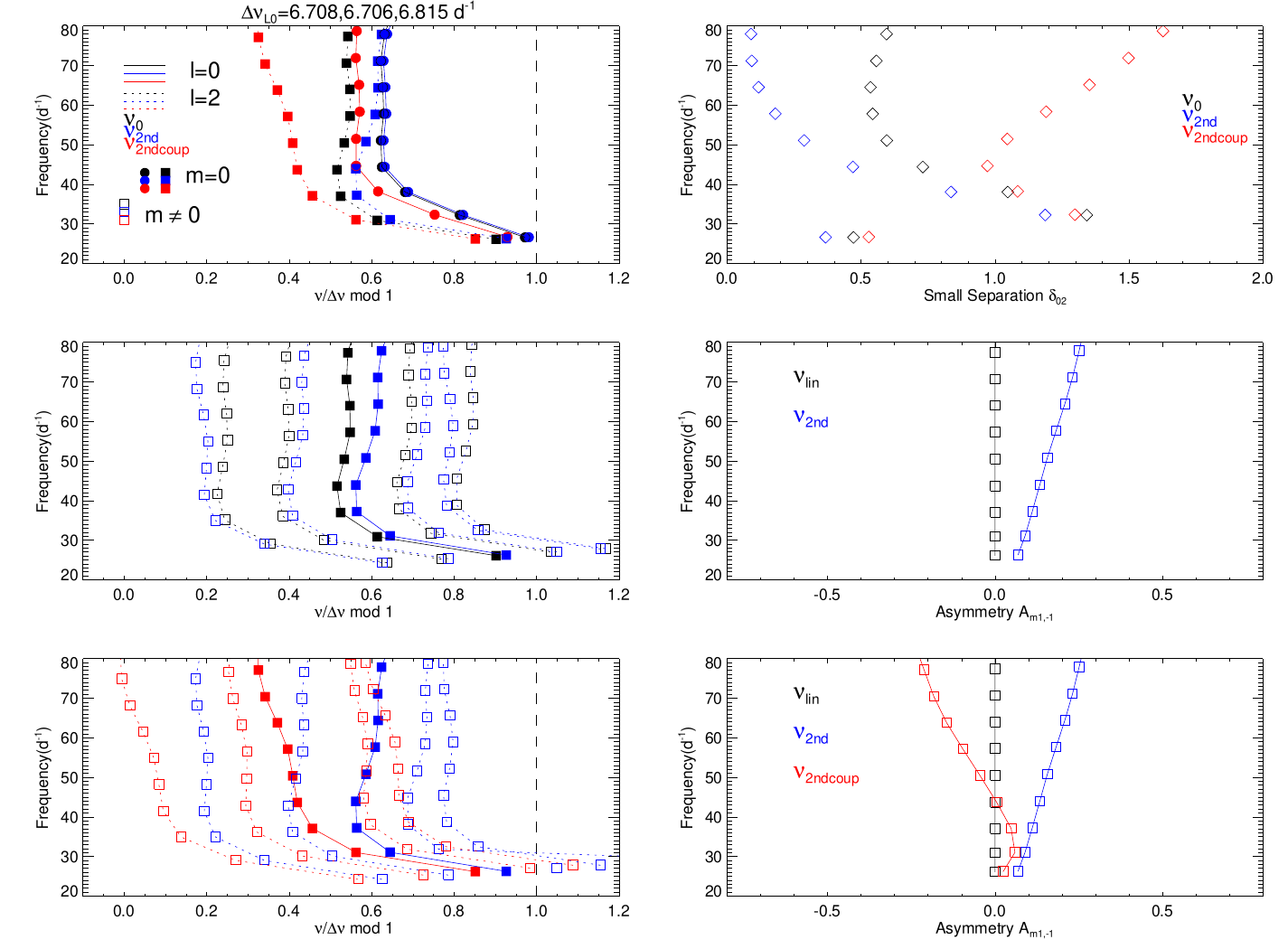}
    \caption{Second-order rotational effect on the frequencies for a $M=1.80$\,M$_\odot$, $Z=0.02$, $v_{\rm eq}=80$\,km~s$^{-1}$ model.   $\nu_0$, $\nu_{\rm lin}$, $\nu_{\rm 2nd}$, $\nu_{\rm 2ndcoup}$ denote frequencies with no rotational correction, linear correction, second-order correction, second-order correction with near-degeneracy coupling, respectively. \textbf{Upper:} Echelle diagram of $l=0,m=0$ and $l=2,m=0$ modes, connected by solid and dotted lines, respectively. Black, blue and red symbols represent the $\nu_{\rm lin}$, $\nu_{\rm 2nd}$, $\nu_{\rm 2ndcoup}$, respectively. Note that the $m=0$ modes are the same in the no-rotation $\nu_0$ and linear treatment $\nu_{\rm lin}$. The right panel shows the variation of small separation $\delta_{02}$ (in d$^{-1}$) as a function of frequency. \textbf{Middle:} Rotational splitting of $l=2$ modes. The linear splitting $\nu_{\rm lin}$ and second-order splitting $\nu_{\rm 2nd}$ are shown. The right panel shows the asymmetry parameter $A_{\rm m1,-1}$. \textbf{Lower:}  Similar to the middle plot but for $\nu_{\rm 2nd}$ and $\nu_{\rm 2ndcoup}$. The corresponding asymmetry parameters are shown in the right panel. Note that the $A_{m1,-1}$ from $\nu_{\rm 2ndcoup}$ (red symbols) change the sign for higher-order modes. }
    \label{fig:example_figure}
\end{figure*}

The above results can also be observed in Fig.\,A1, where we employ a model with $M=1.8$\,M$_\odot$, $v_{\rm eq}=80$\,km~s$^{-1}$. We demonstrate with the echelle diagram to better show the changes of frequency regularities.
We measure the $\Delta \nu$ by performing a linear fit to radial modes ranging from $n=5$ to $8$ \citep{mur23}$:\omega(l=0) =(n+\epsilon)\Delta\nu$. 
 
In the echelle diagram in Fig.\,A1, we use three different large frequency separations calculated by performing the above fitting to $\omega_0$, $\omega_{2nd}$, and $\omega_{\rm 2ndcoup}$ (denoted by $\nu_{0}$, $\nu_{\rm 2nd}$ and $\nu_{\rm 2ndcoup}$ in the figure). This yields $\Delta \nu=6.708, 6.706, 6.815$ d$^{-1}$, respectively. 

In the upper left panel, for the radial ridges $l=0$, we can see that compared to no-rotation ridge based on $\omega_0$,  the $\omega_{\rm 2ndcoup}$ radial ridge shifts to the left significantly (red vs. black). A similar shift occurs for $l=2, m=0$ ridge, but it is more pronounced.

In the middle left panel, we display only the $l=2$ ridges. The black and blue symbols represent $\omega_{\rm lin}$ and $\omega_{\rm 2nd}$ respectively. Rotation splits the ridges to five different ridges, with $m=-2, -1 , 0, 1, 2$ from left right, and open symbols are for $m \neq 0$ modes. It is evident that the black ridges are symmetric and the blue ridges are asymmetric, with larger spacing for the retrograde modes $m=-2, -1$. The middle right panel shows the corresponding asymmetry parameter $A_{m1,-1}$, noting that $A=0$ for $\omega_{\rm lin}$ and $A>0$ for $\omega_{\rm 2nd}$. The asymmetry parameter is larger for larger $n$ modes. 

In the lower left panel, we again show the $l=2$ ridges only but now compare $\omega_{\rm 2nd}$ (blue, same we in middle left panel) with $\omega_{\rm 2ndcoup}$ (red).
Focusing on the $l=2, m=0$ (filled squares), when near-degeneracy mode coupling is included, the $\omega_{\rm 2ndcoup}$ ridge in red shifts to the left substantially. Note that the near-degeneracy effect generally does not alter the $m \neq 0$ mode frequencies (open symbols). The apparent shift is due to the use of different large separations in the echelle plot.

As shown in the bottom panel, the $\omega_{\rm 2nd}$ all show positive asymmetry in the $l=2$ multiplets (blue squared), when near-degeneracy effect is included ($\omega_{\rm 2ndcoup}$), the asymmetry $A_{m1,-1}$ becomes negative for p mode with $n>4$.

The frequency difference between the closest $l=0$ and $l=2$ modes ($\delta_{02}$) decreases as $n$ increases, suggesting a stronger near-degeneracy effect for higher-frequency radial modes. The frequency difference between $l=1$ and $l=3$ modes is much larger compared to the $l=0$ and $l=2$ modes. For such modes, second-order non-spherical correction dominates over near-degeneracy effects.


Asteroseismic modeling has become feasible for certain young $\delta$\,Sct stars, after the discovery of this class of nice $\delta$~Sct stars with regular patterns. The second-order effect and near-degeneracy is typically not included, which renders significant differences in the result. Briefly, the overall effect of rotation, with second-order perturbation and near-degeneracy, is to shift the $m=0$ modes to higher frequency, increase the radial-mode ridge curvature parameter $d_1$, enlarge the large frequency separation $\Delta\nu$,  and reduce the echelle ridge-offset parameter $\epsilon$.
In the example shown in Fig.\,A1, transiting from no rotation ($\nu_0$) to second-order rotational treatment with near-degeneracy coupling ($\nu_{\rm 2ndcoup}$), $\Delta \nu$ is increased by about $0.1$ d$^{-1}$ (from $6.689$ to $6.800$ d$^{-1}$). The $d_1$ parameter increases from $0.46$ to $0.48$, and the corresponding 
$\epsilon$ shifts from $1.624$ to $1.558$. We defer the detailed asteroseismic modeling results to a future paper.






\begin{table*}
	\centering
	\caption{Summary of stellar parameters and splittings of seven $\delta$\,Sct or $\beta$\,Cep stars}
	\label{tab:exple_table1}
	\begin{tabular}{lccccccr} 
	\hline
		Name  & Reference & Type &  &  & \\
		 		\hline
      KIC~10080943 & \citet{sch15}        & $\delta$\,Sct/Binary    &     &   &   \\
\hline
      & M (M$_\odot$)       & R (R$_\odot$)     &$v\sin i $ (km~s$^{-1}$)        &$T_{\rm eff}$ (K)        \\
        star a&   $2.0\pm 0.1$   &   $2.9\pm 0.1$  &    $19.0\pm 1.3$&        $7100\pm 200$\\
 star b&   $ 1.9\pm 0.1$   &   $2.1\pm 0.2$  &    $14.4\pm 1.4$&        $7480^{+180}_{-200}$\\
     \cline{2-5}
&	Frequency $(m=0)$ & $\Delta f -$  & $\Delta f+$ & Asymmetry ($A_{m1,-1}$)  & \\
 &         $12.76334$    & 0.13362(2)& 0.127202(5)   &  0.02460(8)     \\
 \cline{2-5}
 &         $15.10724$   &0.12312(1)     &0.119406(13)     &  0.01531(8)      \\ 
                 \cline{2-5}
 &         $19.51028$   & 0.13448(4)     & 0.12860(2)   &  0.02235(18)   & \\
 \hline
     KIC~9244992 &  \citet{sai15}      &      $\delta$\,Sct &        &        \\
\hline
      & M (M$_\odot$)       & R (R$_\odot$)     &$v\sin i$ (km~s$^{-1}$)        &$T_{\rm eff}$ (K)         \\
        &   $1.45$   &   $2.03$  &  $<6$  &        $7000^{+300}_{-100}$\\
   \cline{2-5}
&Frequency $(m=0)$ & $\Delta f -$ & $\Delta f+$ & Asymmetry ($A_{m1,-1}$)  & \\
                 \cline{2-5}
       &          12.906366   &      0.013838(9)     &      0.013692(7)   &      0.0053(4)     \\
                 \cline{2-5}
       &          13.995924   &       0.015763(73)      &      0.015122(71)   &       0.0207(33)     \\
                 \cline{2-5}
       &          14.061786   &       0.018019(77)     &      0.016755(56)   &       0.0364(27)     \\
                 \cline{2-5}
       &          14.806054   &        0.014411(105)      &      0.014412(69)   &    0.000(4)     \\
                 \cline{2-5}
       &          15.604826   &           0.014735(112)      &      0.014549(13)   &      0.0063(59)     \\
                 \cline{2-5}
       &          16.404983   &       0.015017(23)      &      0.0150795(155)   &     -0.0021(9)     \\
    \hline
     KIC~11145123 &  \citet{kur14}, Table 2      & $\delta$\,Sct     &        &        \\
\hline
      Partially shown & M (M$_\odot$)       & R (R$_\odot$)     & P$_{\rm rot}$(d)        &$T_{\rm eff}$ (K)         \\
        &   $\approx 1.46$   &   $\approx 2.24$  & $\sim 100$   &        $8050\pm200$\\
                 \cline{2-5}
	&Frequency $(m=0)$ & $\Delta f -$ & $\Delta f+$ & Asymmetry ($A_{m1,-1}$)  & \\
       &   $18.3660001$   &0.0101696(14)     & 0.0101209(14)   &  0.00240(96)     \\
                \cline{2-5}
      &   $22.0018915$   &0.00856(3)      & 0.00853(3)   &   0.0017(24)     \\
                 \cline{2-5}
      &   $23.5160925$   &0.00990(19)     & 0.00976(19)   &   0.0069(136)     \\
                 \cline{2-5}
       &   $23.8185035$   &0.010213(13)      & 0.010179(91)   &   0.0017(78)    \\
                 \cline{2-5}
       &   $24.4192854$   &0.009887(6)     & 0.00988(8)   &  0.00035(495)    \\
\hline
     HD~192575 & \citet{bur23}        & $\beta$\,Cep     &        &        \\
 \cline{2-5} 
      & M (M$_\odot$)       & R (R$_\odot$)     &$v\sin i$ (km~s$^{-1}$)        &$T_{\rm eff}$ (K)         \\
        &   $12.0\pm 1.5$   &   $\sim 9.1^{+0.8}_{1.7}$  &$27^{+6}_{-8}$    &     $23900\pm 900$\\
 \cline{2-5}
	&Frequency $(m=0)$ & $\Delta f -$ & $\Delta f+$ & Asymmetry ($A_{m1,-1}$)  & \\
       &   $6.696791(5)$   &0.170217(7)     & 0.171786(6)   &   0.004589(3)     \\	
 \cline{2-5}
       &   $6.46341(1)$   &0.16731(2)    & 0.17058(2)   &   0.00968(8)     \\	
		\hline
     HD~129929 & \citet{aer04}        &      $\beta$\,Cep&        &        \\
\hline
      & M (M$_\odot$)       & R (R$_\odot$)     &$v\sin i$ (km~s$^{-1}$)        &$T_{\rm eff}$ (K)         \\
        &   $9.0-9.5$   &   $-$  &$ \le 13$    &     $\sim 23990$\\	
 \cline{2-5}
	&Frequency $(m=0)$ & $\Delta f -$ & $\Delta f+$ & Asymmetry ($A_{m1,-1}$)  & \\
       &   $6.97831(1)$   &0.012133(30)     & 0.012126(30)   &   0.00029(58)     \\
\hline
     $\nu$ Eridani & \citet{han04}        &$\beta$\,Cep      &        &        \\
\hline
      & M (M$_\odot$)       & R (R$_\odot$)     &$v\sin i$ (km~s$^{-1}$)        &$T_{\rm eff}$ (K)        \\
        &   $\sim 9.86$   &   $\sim 6.28$  &$ \approx 20$    &     $\sim 22180 \pm 100$\\	
 \cline{2-5}
	&Frequency $(m=0)$ & $\Delta f -$ & $\Delta f+$ & Asymmetry ($A_{m1,-1}$)  & \\
       &   $5.63716(2)$   &0.01710(3)      & 0.01677(3)   &   0.0097(13)     \\	
\hline
     $\theta$ Oph & \citet{bri07}        &$\beta$\,Cep/triple      &        &        \\
\hline
    & M (M$_\odot$)       & R (R$_\odot$)     &$v_{\rm eq}$ (km~s$^{-1}$)        &$T_{\rm eff}$ (K)        \\
       &   $8.2\pm 0.3$   &   $5.0\pm 0.4$  &$ 29\pm 7$    &     $22 260 \pm 280$\\	
 \cline{2-5}
	&Frequency $(m=0)$ & $\Delta f -$ & $\Delta f+$ & Asymmetry ($A_{m1,-1}$)  & \\	
       &   $7.8742(20)$   &0.1083(30)     & 0.0992(30)   &   0.044(20)      \\
\hline
	\end{tabular}
\end{table*}



\bibliographystyle{mnras}
\bibliography{asymref}

\bsp	
\label{lastpage}
\end{document}